\documentclass[namedreferences]{SolarPhysics}
\usepackage[optionalrh]{spr-sola-addons} 
\usepackage{solaheader}

\newcommand{\new}{}

\usepackage{graphicx}        
\usepackage{amssymb}         
\usepackage{mathrsfs}
\usepackage[usenames,dvipsnames]{xcolor}
\usepackage{url}             

\usepackage{epstopdf}
\usepackage{amsmath}

\newcommand{\half}{{\textstyle\frac{1}{2}}}

\newcommand{\re}{\mathop{\rm Re}\nolimits}
\newcommand{\im}{\mathop{\rm Im}\nolimits}

\newcommand{\sech}{\mathop{\rm sech}\nolimits}

\newcommand{\rd}{\mathrm{d}}

\newcommand{\rmi}{\mathrm{i}}
\newcommand{\rme}{\mathrm{e}}

\newcommand{\pderiv}[2]{\frac{\partial#1}{\partial#2}}
\newcommand{\pderivd}[2]{\frac{\partial^2#1}{\partial#2^2}}

\newcommand{\deriv}[2]{\frac{\rd#1}{\rd#2}}

  \renewcommand{\le}{\leqslant}

\newcommand{\aap}{\textit{Astron. Astrophys.}}

\newcommand{\apj}{\textit{Astrophys. J.}}
\newcommand{\apjs}{\textit{Astrophys. J. Suppl.}}

\newcommand{\nat}{\textit{Nature}}

\newcommand{\solphys}{\textit{Solar Phys.}}

\newcommand{\ssr}{\textit{Space Sci. Rev.}}

\begin{document}
\begin{opening}

\title{\new Alfv\'en Reflection and Reverberation in the Solar Atmosphere}

\author{P. S. \surname{Cally}}
\institute{Monash Centre for Astrophysics and School of
Mathematical Sciences, Monash University, Victoria 3800, Australia \email{paul.cally@monash.edu}}

\runningauthor{P.S. Cally}

\runningtitle{\new Alfv\'en Reflection and Reverberation}

\begin{abstract}
Magneto-atmospheres with Alfv\'en speed [$a$] that increases monotonically with height are often used to model the solar atmosphere, at least out to several solar radii. A common example involves uniform vertical or inclined magnetic field in an isothermal atmosphere, for which the Alfv\'en speed is exponential. We address the issue of internal reflection in such atmospheres, both for time-harmonic and for transient waves. It is found that a mathematical boundary condition may be devised that corresponds to perfect absorption at infinity, and, using this, that many atmospheres where $a(x)$ is analytic and unbounded present no internal reflection of harmonic Alfv\'en waves. However, except for certain special cases, such solutions are accompanied by a wake, which may be thought of as a kind of reflection. For the initial-value problem where a harmonic source is suddenly switched on (and optionally off), there is also an associated transient that normally decays with time as $\mathcal{O}(t^{-1})$ or $\mathcal{O}(t^{-1}\ln t)$, depending on the phase of the driver. Unlike the steady-state harmonic solutions, the transient does reflect weakly. Alfv\'en waves in the solar corona driven by  a finite-duration train of $p$-modes are expected to leave such transients.
\end{abstract}



\end{opening}

\section{Introduction}
Alfv\'en waves in a stratified atmosphere are governed by the standard linear wave equation
\begin{equation}
\pderivd{\xi}{t}=a^2 \pderivd{\xi}{x}\, ,  
\label{alfeqn}
\end{equation}
where $\xi$ is the plasma displacement (which is transverse to both the magnetic field and the direction of inhomogeneity $x$), and $a=a(x)=|B_x|/\sqrt{\rho}$ is the Alfv\'en speed, or more properly the Alfv\'en velocity component in the $x$-direction (the magnetic permeability is scaled to unity throughout). In large-scale open field regions of the Sun's atmosphere, the Alfv\'en speed increases monotonically with height due to the decreasing density [$\rho$], with geometric diminution of magnetic-field strength playing a secondary role. Eventually, $a$ reaches a maximum at several $R_\odot$ and decreases with distance thereafter {\new\cite{AnSueMoo90aa}}. By this stage though, solar-wind flows have become important and Equation (\ref{alfeqn}) must be modified to take these into account {\new\cite{Vel93aa}}.

However, our focus in this article lies below such heights, and with wave periods of minutes rather than the hours common in the heliosphere. We simply ask, \emph{What is the nature of Alfv\'enic solutions of the simple Equation (\ref{alfeqn})?} We are all very familiar with the basic wave equation, treated at length in every text book on partial differential equations (PDEs). However, the nonuniform Alfv\'en speed introduces some features that are perhaps less well known, and that have significance for the nature of Alfv\'enic oscillations now seen in the solar corona \cite{TomMcIKei07aa,McIde-Car11aa}.

We shall address a range of analytic Alfv\'en speed profiles, but the most basic is 
\begin{equation}
a=a_1(x)=a_0\, \rme^{x/2h}\,,
\label{Alf1}
\end{equation}
which pertains to a uniform magnetic field and an isothermal density stratification (scale height $h$). This model has been much used as a fundamental representation of Alfv\'en waves in stellar atmospheres at least since \inlinecite{Fer54aa}. \inlinecite{FerPlu58aa} noted the exact solution
\begin{equation}
\xi = \left[A_1 \, J_0\left(2\,\frac{\omega h}{a_0}\rme^{-x/2h}\right) + A_2\,Y_0\left(2\,\frac{\omega h}{a_0}\rme^{-x/2h}\right)\right] \rme^{-\rmi\omega t}
\label{ferraro}
\end{equation}
for a wave of single frequency [$\omega$] in terms of Bessel functions of the first and second kind of order zero, with $A_1$ and $A_2$ arbitrary constants. They then dropped the $Y_0$ solution on the grounds that the velocity perturbation does not vanish as $x\to\infty$ (\emph{i.e.} as $\rho\to0$). We term this the regularity boundary condition. Many subsequent studies have adopted the same approach  \cite[for example]{AnMusMoo89aa}. This is problematic though; it imposes a perfectly reflecting boundary at infinity thereby setting up a standing wave $J_0\left(2\rme^{-x/2h}\,{\omega h/a_0}\right)$, which may not be desirable or realistic. The regularity boundary condition is unnecessary from an energy point of view too. Despite the $Y_0$ solution being unbounded, $\mathcal{O}(x)$ in fact, the kinetic-energy density [$\half\rho\omega^2|\xi|^2$] vanishes at infinity, and the magnetic energy density $\half |b|^2$ is finite there, where $b$ is the magnetic field perturbation.\footnote{A quadratic wave-energy equation is easily constructed from the linearized momentum and induction equations:
$
\partial\mathcal{E}/\partial{t} + \partial{\mathcal{F}}/\partial{x}=0\,,
$
where $\mathcal{E}=\half\rho v^2 + \half b^2$ is the energy density, $\mathcal{F}=-B\,b\,v$ is the wave-energy flux, $v=\partial\xi/\partial t$ is the plasma velocity, and $b=-B\, \partial\xi/\partial x$ is the magnetic field perturbation. Energy fluxes may be attributed to solutions of the wave equation using the formula for $\mathcal{F}$, and hence reflection coefficients may be calculated if these solutions can be split into upward- and downward-propagating parts. If $\xi$ and $b$ are being modelled as complex quantities, then we should write $\mathcal{E}=\half\rho |v|^2 + \half |b|^2$ and $\mathcal{F}=-B\,\re(b\,v^*)$.} The $\xi=\mathcal{O}(x)$ behaviour is just the linear flapping of rigid field lines to be expected physically as $a\to\infty$ (see the animation attached to \opencite{CalHan11aa}).

Although mathematically we are at liberty to impose the regularity boundary condition, it is often more convenient in theoretical modelling to allow waves to escape at the top so as not to confuse matters with downward travelling waves reflected from an unphysical ``infinity''. The boundary condition at infinity is particularly important in the exponential model since the Alfv\'en travel time [$\tau=2h/a(x)$] from any point $x$ to infinity is finite. To place this in a solar coronal context, assuming a 2 G magnetic field, a base density of $10^{-12}$ $\rm kg\,m^{-3}$, and a density scale height of 20 Mm, the travel time to infinity in an exponential atmosphere is around 450 seconds, comparable to the period of the waves that we are considering.

Despite Alfv\'en waves undoubtedly reflecting from sharp features such as the chromosphere-corona transition region \cite{cravan05aa,HanCal12aa}, or returning along closed loops, there is reason to believe that waves in the open-field corona are preferentially outgoing. This is partly because some fraction of their energy irrevocably escapes into the solar wind, and partly because of damping by such mechanisms as turbulent cascade to the ion-cyclotron scale (see for example \opencite{MarVocTu03aa}, \opencite{Hol06aa}), neither of which we choose to model here. So, a radiation condition is a way of avoiding such complications when exploring propagation far below where they are situated. 

The wave-energy flux associated with the general solution (\ref{ferraro}) is
\begin{equation}
\begin{split}
\mathcal{F} &= \frac{\omega\,B^2}{4\,h\,\pi} \left( |A_1+\rmi\,A_2|^2 -|A_1-\rmi\,A_2|^2\right)\\[4pt]
&= \frac{\omega\,B^2}{h\,\pi} \left( |\Xi_2|^2 -|\Xi_1|^2\right)
\end{split}
\label{JYflux}
\end{equation}
where we define $\Xi_1=\half(A_1-\rmi\,A_2)$ and $\Xi_2=\half(A_1+\rmi\,A_2)$.  This shows that the alternate but equivalent representation
\begin{equation}
\xi = \left[\Xi_1 \, H_0^{(1)}\left(2\,\frac{\omega h}{a_0}\rme^{-x/2h}\right) + \Xi_2\,H_0^{(2)}\left(2\,\frac{\omega h}{a_0}\rme^{-x/2h}\right)\right] \rme^{-\rmi\omega t}\,,
\label{ferraroH}
\end{equation}
separates the solution into downgoing and upgoing parts respectively. The solution with $\Xi_1=0$,
\begin{equation}
\xi_\uparrow =  \, H_0^{(2)}\left(2\,\frac{\omega h}{a_0}\rme^{-x/2h}\right)\,,
\label{Hank2}
\end{equation}
was used by \inlinecite{SchCalBel84aa} to represent an outward-travelling wave (the harmonic temporal dependence is omitted from now on, but is implied). This Hankel function (Bessel function of the third kind) is well known to asymptotically reduce to a complex exponential $H_0^{(2)}(r)\sim\sqrt{2/(\pi r)}\,\exp[-\rmi(r-\pi/4)]$ for large argument $r\to\infty$ ($x\to-\infty$) \cite[formulae 10.17.5-6]{DLMF}
(\inlinecite{Hol78aa} also uses a Hankel-function solution to represent a radiation boundary condition, but assumes that this is valid only ``sufficiently far from the Sun that the eikonal approximation is valid and no more wave reflections are expected''.)

\inlinecite{HanCal12aa} point out that the Alfv\'en wave equation with exponential Alfv\'en speed $a_1(x)$ is isomorphic to the uniform axisymmetric two-dimensional (2D) wave equation with unit wave speed
\begin{equation}
\pderivd{\xi}{t} = \frac{1}{r} \pderiv{}{r}\left( r \pderiv{\xi}{r} \right)
\label{axisym}
\end{equation}
under the transformation $r = 2 h/a$. Here the axis $r=0$ corresponds to $x=+\infty$ in the Alfv\'en wave problem. The textbook solution $J_0(\omega r)=\half[H_0^{(1)}(\omega r)+H_0^{(2)}(\omega r)]$ effectively imposes perfect reflection on the axis, resulting in a standing wave. Placing a harmonic source there instead yields the $H_0^{(1)}(\omega r)$ solution (\opencite{CouHil62aa}, Ch.~III, Section 3.2; \opencite{Whi74aa}, Section 7.4) whilst the time reverse corresponds to a perfect absorber and yields $H_0^{(2)}(\omega r)$. There is nothing reflective in this system. But neither are the waves represented by simple d'Alembert-like solutions. As is well known and understood \cite[pp.~208--210]{Had23aa,CouHil62aa}, solutions of the wave equation for uniform media in even-dimensional spaces do not obey Huygens' principle. Any propagating disturbance trails a wake, or ``reverberation''; to quote \inlinecite{Joh82aa}, ``Disturbances propagate with finite speed but after having reached a point never die out completely in a finite time at that point, like the waves arising from a stone dropped into water''. But even more restrictedly, only in one and three spatial dimensions do \emph{spherical waves} propagate ``relatively undistorted'' \cite[Ch.~VI, Section 18]{CouHil62aa}. The 2D case of interest here necessarily exhibits a wake.

As noted by \inlinecite{HanCal12aa}
\begin{equation}
\begin{split}
H_0^{(1,2)}(r) &=
 \frac{2}{\pi}\int_0^1\frac{\rme^{\pm \rmi\,u\,r}}{\sqrt{1-u^2}} \,\rd u  \mp  \frac{2i}{\pi}\,\int_0^\infty \rme^{-r\sinh\tau}\rd\tau \\[8pt]
 &= \alpha_\pm(r) \mp \rmi\, \beta(r)
 \, .
 \end{split}
 \label{Hankel2}
\end{equation}
Clearly, $\alpha_+$ corresponds to propagation away from $r=0$ (backward in $x$), $\alpha_-$ represents propagation towards $r=0$, and $\beta$ is the reverberation. We see that the propagation part of $\xi(r)$, \emph{i.e.} $\alpha_\pm(\omega r)$, consists of a superposition of all wavenumber components between 0 and $\omega$. Of course, $\beta$ may also be represented as a Fourier integral
\begin{equation}
\beta(r) =  \frac{2}{\pi}\int_0^1 \frac{\sin u r\, \rd u}{\sqrt{1-u^2}} +  \frac{2}{\pi} \int_1^\infty \frac{\cos u r\, \rd u}{\sqrt{u^2-1}}\,,
\label{reverbstand}
\end{equation}
showing that the reverberation can be represented as a superposition of standing waves, in agreement with the {\new characteristics-based} findings of \inlinecite{HolIse07aa}. In terms of Bessel and Struve functions, $\alpha_\pm(r)=J_0(r)\pm \rmi\,\mathbf{H}_0(r)$ and $\beta(r)=\mathbf{H}_0(r)-Y_0(r)$; note that $\alpha_\pm$ is regular at $r=0$ \cite[Figure 10]{HanCal12aa}. {\new For small $r$, the reverberation [$\beta$] exhibits a logarithmic singularity inherited from $Y_0$, $\beta\sim -2\,\pi^{-1} \ln r$. This will be seen again later (Sections \ref{switchon} and \ref{switchonoff}) in the asymptotic form of the transients.}

The fact that Alfv\'en waves in the inner corona appear to be preferentially outgoing suggests that $H_0^{(2)}(\omega r)$ is a better model of solar atmospheric Alfv\'en wave propagation than is $J_0(\omega r)$, and an indication that strong absorption occurs before the waves can reflect.  In any case, we are mathematically at liberty to impose a radiation condition at large $x$, and it is certainly convenient to do so in theoretical investigations of outward wave propagation. This article, in part, addresses the consequences of this choice, both for the steady-state harmonic wave problem and for the initial-value problem.

Indeed, it is very common to seek to impose a radiation condition above an exponential or similar atmosphere by simply appending a uniform plasma above some (large) height {\new\cite{Har29aa}}. However, it will be shown that this is not a good choice in general because the discontinuity in Alfv\'en speed gradient is itself highly reflective. The artifice therefore decides the issue rather than illuminates it.

Another interesting suggestion for how to avoid the difficulties presented by a finite travel time to infinity and perfect reflection there is to retain the displacement current in the wave equations \cite{Ler83aa,TsaSteKop09aa}. Then the Alfv\'en speed is limited by the speed of light, and the Alfv\'en wave ultimately couples to an outgoing electromagnetic wave. (It should be noted though that \citeauthor{Ler83aa} did not mean to imply that this process actually occurs in the solar atmosphere -- see the last sentence of his article -- the device was merely introduced as a mathematical way around the troublesome infinity.) Unfortunately, we shall see in Section \ref{flat} that this model is almost totally reflective for any realistic frequency and density scale height, and so does not fulfil our requirement for optimal transmission.

Reflection of harmonic waves will be addressed from several standpoints and for a variety of atmospheres in Section \ref{sec:reflect}, showing that the reflectivity of an atmosphere is governed by its smoothness, or lack thereof. Discontinuity in any derivative of the Alfv\'en speed results in a reflection coefficient that depends algebraically on the wavenumber. On the other hand, an infinitely smooth ($\mathscr{C}^\infty$) function suffers only exponentially small reflection at worst. Then we address the initial-value problem, in the exponential atmosphere, with specific focus on the decay of transients, both for the radiation and regularity boundary conditions. Finally, we draw some general conclusions about the relationship between reflectivity, wakes, and transients and how they are determined by the form of $a(x)$.


\section{Reflection} \label{sec:reflect}
\subsection{Power Law Alfv\'en Profiles and the Prevalence of Wakes}
One might expect that any nonuniform wave-speed profile gives rise to reflection. This is not the case though. For example, it is known that the profile $a=x^2$ admits relatively undistorted exact solutions of the form $\xi=x\,U(t\pm x^{-1})$ that represent unidirectional propagation \cite{DidPelSoo08aa}. 

For other power laws $a=a_0 x^{(n-1)/(n-2)}$ ($n\ne2$, with $x>0$ if required) the wave equation transforms to
\begin{equation}
\pderivd{\xi}{t}= \pderivd{\xi}{r} + \frac{n-1}{r}\,\pderiv{\xi}{r}
\label{apower}
\end{equation}
under the change of variables $r=(n-2)\,a_0^{-1}x^{-1/(n-2)}$. The profile $a(x)$ is monotonic increasing if $n>2$ or $n<1$, with the former case yielding finite Alfv\'en travel time to infinity, $\int^\infty a(x)^{-1}\,\rd x <\infty$. For $n\le1$ the Alfv\'en exponent $(n-1)/(n-2) \in[0,1)$ and the travel time to $x=+\infty$ is infinite, so the issue of reflection from infinity does not arise in any initial-value calculation. For positive integer $n$ (although we have no need to restrict it thus) Equation (\ref{apower}) is the spherical wave equation in $n$ dimensions. In one dimension the well-known d'Alembert solution is $\xi(x,t)=U(t\pm r)$ for arbitrary shape function $U$ (this is strictly undistorted), and the analogue for three dimensions is $\xi(r,t) = r^{-1}U(t\pm r)$ (relatively undistorted). 
The $n=3$ case corresponds to the $a=x^2$ profile mentioned above.

Exact harmonic solutions of Equation (\ref{apower}) may be represented in terms of Hankel functions,
\begin{equation}
\xi = r^\mu H_{|\mu|}^{(1,2)}(\omega r)\,\rme^{-\rmi\omega t}\,,
\label{apowerHank2}
\end{equation}
where $\mu=1-(n/2)$. These solutions reduce to elementary functions without branch points only when $n$ is an odd integer (spherical Bessel functions). For $n=1$ ($a=1$, arbitrarily scaling $a_0$ to unity), $\xi=\rme^{\rmi\omega(\pm r-t)}=\rme^{\rmi\omega(\pm x-t)}$, from which the D'Alembert solutions clearly may be constructed by Fourier composition. For $n=3$ ($a=x^2$), $\xi=r^{-1}\rme^{\rmi\omega(\pm r-t)}= x\, \rme^{\rmi\omega(\pm x^{-1}-t)}$, which again obviously leads to $\xi=x\,U(t\pm x^{-1})$ for arbitrary $U$. 

For $n=5$ ($a=x^{4/3}$), $\xi=r^{-3}(\omega r\mp i)e^{- i \omega(t\pm r)}$ with $r=3x^{-1/3}$, showing that solutions of the form $\xi= r^{-3} U_\pm(r\pm t) - r^{-2}U_\pm'(r\pm t)$ for arbitrary differentiable $U_\pm$ may be synthesized. Initial displacement specifies $U_++U_-$ and initial velocity gives $U_+-U_-$. If $\xi(r,0)$ and $\xi_t(r,0)$ have joint compact support, $\xi(r,t)$ will continue to do so, with the interval on which $\xi\ne0$ simply translating at unit speed in each direction. No wake is left, although the shape of the pulse changes. Higher odd-$n$ cases behave similarly, with progressively higher derivatives of $U$ involved. In fact, $\xi=(\partial^2/\partial t^2-\nabla^2)^{(n-1)/2}\phi(t\pm r)$ is a solution for arbitrary $\phi$, where $\nabla^2$ is the $n$-dimensional spherical Laplacian, \emph{i.e.}, the right-hand side of Equation (\ref{apower}) \cite[Ch. VI, Section 13.4]{CouHil62aa}. 

For all other real powers, and for the exponential profile that corresponds to $n=2$ in Equation (\ref{apower}), radiating solutions contain logarithmic or fractional power singularities at $r=0$. These branch points give rise to continuous spectra in the initial value problem, and hence to transients that decay algebraically with time. Simple ``relatively undistorted'' translating solutions exist only for $n=1$ and 3, and compact support is simply translated for other odd $n$. Otherwise, propagating wave packets trail wakes. This is the rule rather than the exception in inhomogeneous media. But do we get reflection other than that implicit in the wake?


\subsection{WKB Perspective}

Within the WKB approximation \cite{BenOrs78aa,Gou07aa}, one seeks asymptotic solutions of $a^2 \xi_{xx}+\omega^2 \xi=0$ that take the form 
\begin{equation}
\xi=\exp\left[\omega\sum_{m=0}^\infty\omega^{-m}S_n(x)\right].   \label{wkb}
\end{equation}
Truncating at $m=1$ (the physical optics approximation), the general solution for arbitrary $a(x)$ is given by
\begin{equation}
\xi \sim a(x)^{1/2}  \left[A\,\rme^{\rmi \,\varphi(x)} + B\,\rme^{-\rmi\,\varphi(x)}\right]\,, \qquad \mbox{$\omega\to\infty$, \emph{i.e.} $k\to\infty$,}
\end{equation}
formally valid for $k\gg |d\ln k/dx|$,
where $k(x)=\omega/a(x)$, $\varphi(x)=\int^x k(x)\,\rd x=-i\,S_0$, and $A$ and $B$ are arbitrary constants representing the amplitudes of the upward and downward waves respectively. The two waves are completely decoupled, meaning that there is no reflection. Coupling is not recovered by going to higher order in $\omega$. The reason for this is that the original assumption (\ref{wkb}) does not admit such coupled solutions.

However, this all breaks down where $k^{-1}|d\ln k/dx| \not\ll 1$, and particularly at points of discontinuity in $a$ or $a'$. For discontinuous Alfv\'en speed, the wave-energy reflection coefficient results from the $m=0$ term and is given by $\mathcal{R}=(a_+ -a_-)^2/(a_++a_-)^2$ to leading order, independent of frequency, where $a_\pm$ are the Alfv\'en speeds on either side of the discontinuity. For discontinuous Alfv\'en speed slope the effect enters at $m=1$ and yields $\mathcal{R}=(\Delta a')^2/(16\omega^2+(\Delta a')^2)$, where $\Delta a'=a_+'-a_-'$ is the jump in slope. Notice that reflection vanishes as $\omega\to\infty$ in this case. \inlinecite{Vel93aa} makes the interesting point that multiple such discontinuities in scale height can set up resonances because they create (leaky) cavities. Discontinuities in higher order [$d>1$] derivatives are invisible at the physical optics level, appearing only for higher $m=d$ and yielding $\mathcal{O}(\omega^{-2d})$ reflection coefficients. We therefore expect that $\mathcal{R}$ is exponential rather than algebraic for $\mathscr{C}^\infty$ functions $a(x)$ (\emph{i.e.} functions that are continuously differentiable to all orders). This is confirmed in the examples below.

\subsection{Asymptotically Flat Profiles}\label{flat}
Consider several cases where the exponential Alfv\'en profile for $a^2$ is capped at a maximum level $a_u^2=a_0^2/\epsilon$ as $x\to\infty$.

\begin{figure}
\begin{center}
\includegraphics[width=0.666\textwidth]{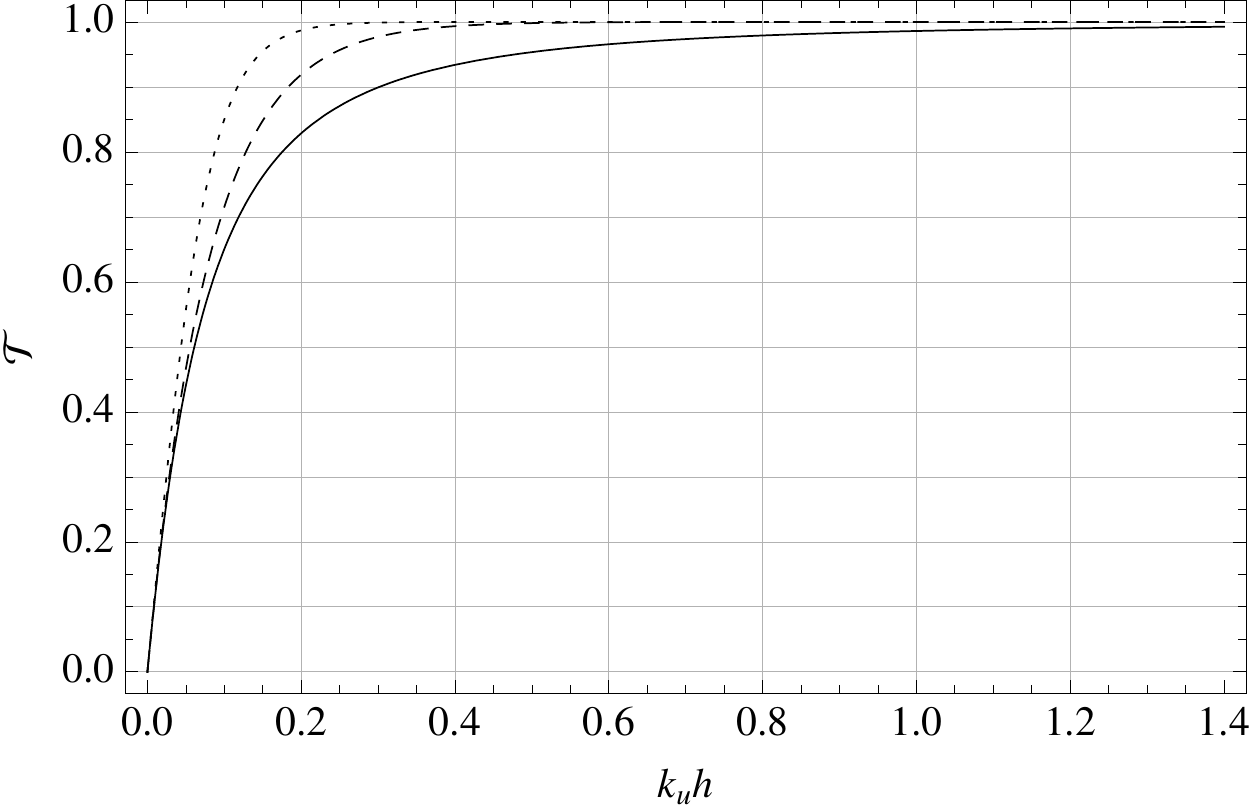}
\caption{Transmission coefficients against $k_uh$ for the three asymptotically flat profiles. $\mathcal{T}_2$ (full curve); $\mathcal{T}_3$ (dashed); $\mathcal{T}_4$ (dotted). All three transmission coefficients are asymptotic to $4\pi k_uh$ as $k_u\to0$.}
\label{fig:T}
\end{center}
\end{figure}

\begin{itemize}
\item{Sharply Capped Exponential:}
For
\begin{equation}
a^2=a_2^2=
\begin{cases}
a_0^2 \, \rme^{x/h} & x<-h\ln\epsilon, \\
a_0^2/\epsilon & x>-h\ln\epsilon.
\end{cases}
\end{equation}
the reflection coefficient is
\begin{equation}
\mathcal{R}_1=\left|\frac{H_0^{(2)}(2 k_uh )+\rmi\,  H_1^{(2)}(2 k_uh )}{H_0^{(1)}(2 k_uh )+\rmi\, 
   H_1^{(1)}(2 k_u h)}\right|^2\sim \frac{1}{64 k_u^2h^2}
   \label{Rsharp}
\end{equation}
as $k_u\to\infty$, where $k_u=\omega/a_u$, in agreement with the matched WKB result above. 

\item{Smoothly Capped Exponential:} With $a^2=a_3^2=a_0^2\,\rme^{x/h}/(1+\epsilon\,\rme^{x/h})$, the harmonic wave equation can be expressed in the form of a Bessel equation,
\begin{equation}
r \deriv{}{r}\left(r \deriv{\xi}{r}\right) + \left(r^2 +4 k_u^2h^2\right)\xi=0\,,
\label{smoothcapeqn}
\end{equation}
where $k_0=\omega/a_0=\epsilon^{-1/2}k_u$ and $r=2k_0h \,\rme^{-x/2h}$. The solution satisfying the radiation condition at $x\to+\infty$ is easily identified using $J_\nu(z)\sim (z/2)^\nu/\Gamma(1+\nu)$ as $z\to0$:
\begin{equation}
\begin{split}
\xi 
   &=2J_{-2\, \rmi \, k_uh}(r)\quad\mbox{since it is $\mathcal{O}(\rme^{i\,k_u x})$ as $x\to+\infty$}\\[4pt]
  &=  H_{-2\, \rmi \, 
   k_uh}^{(1)}\left(r\right)+ H_{-2 \,\rmi \,
   k_uh}^{(2)}\left(r\right) \\[4pt]
 &\sim \sqrt{\frac{2}{\pi\,r}} \bigg\{ \rme^{-\pi\,k_uh} \,\rme^{\rmi(r-\pi/4)} + \rme^{\pi\,k_uh} \,\rme^{-\rmi(r-\pi/4)} \bigg\}
 \quad\text{as $r\to\infty$.}
\end{split}
   \label{smoothcap}
\end{equation}
The reflection coefficient is therefore the squared ratio of the coefficients of the complex exponentials: $\mathcal{R}_3=\rme^{-4 \pi  k_uh}$. As expected since $a(x)$ is analytic, reflection is exponentially small as $k_u\to\infty$. Equation (\ref{smoothcapeqn}) has a regular singular point at $r=0$ ($x=\infty$) with roots of the indicial equation $\mu=\pm 2\,\rmi\,k_uh$. 

\item{Alternate Smoothly Capped Exponential:} With a gentler profile $a^2=a_4^2=a_0^2\,\rme^{x/h}/(1+\epsilon^{1/2}\,\rme^{x/2h})^2$ the wave equation takes the form
\begin{equation}
r \deriv{}{r}\left(r \deriv{\xi}{r}\right) + \left(r +k_uh\right)^2\xi=0\,,
\label{altsmoothcapeqn}
\end{equation}
with $r=2k_0h\, \rme^{-x/2h}$ again. The radiation solution is given in terms of a confluent hypergeometric function,
\begin{equation}
\xi = \rme^{\rmi k_u x -\rmi\,r} \,
   _1F_1\left(\frac{1}{2};1-4 \,\rmi \,k_uh;2\,\rmi\,r\right)\,.
        \label{altsmoothcap}
\end{equation}
Analysis of the large $r$ asymptotics of the ${}_1F_1$ function \cite[formula 13.7.2, where $\mathbf{M}(a,b,z)={}_1F_1(a;b;z)/\Gamma(b)$]{DLMF}
reveals that the reflection coefficient is $\mathcal{R}_4=\rme^{-4 \pi  k_uh}\sech(4 \pi  k_uh)$. Once again, $\mathcal{R}$ decreases exponentially with increasing $k_u$, as expected. In this case, $r=0$ is again a regular singular point, but now the roots of the indicial equation are $\mu=\pm \rmi\,k_u\,h$.
\end{itemize}
The three transmission coefficients $\mathcal{T}=1-\mathcal{R}$ are plotted against dimensionless wavenumber [$k_uh$] in Figure \ref{fig:T}, illustrating the fact that $\mathscr{C}^\infty$ profiles remain essentially fully transparent to much smaller wavenumbers. Lest the reader think that the differences are entirely due to the maximum values of the WKB validity parameter $k^{-1}|\rd\ln k/\rd x|$, \emph{i.e.} $(2k_uh)^{-1}$, $(3\sqrt{3}\,k_uh)^{-1}$, and $(8k_uh)^{-1}$ respectively for the three cases, this does not fully account for the discrepancies.

The three capped models are all totally reflective in the limits $a_u\to\infty$ and $\omega\to0$ both of which correspond to $k_u\to0$, so they represent poor devices for allowing radiation to escape at the top.  {\new This approach led \inlinecite{BelLer81aa} to discount Alfv\'en losses in sunspots.}
As suggested by the figure, a sequence of $\mathscr{C}^\infty$ functions, more and more closely approximating the sharply capped exponential, will of course return the transmission coefficient (\ref{Rsharp}) in the limit.

With $\epsilon=a_0^2/c^2$, where $c$ is the speed of light, the smoothly capped exponential  case $a_3$ corresponds to that of Equation (9) in \inlinecite{Ler83aa}, where the displacement current is retained in the Alfv\'en wave equation and the wave travel speed is limited by $c$. The corresponding reflection coefficient [$\mathcal{R}=\rme^{-4\pi\omega h/c}$] is of course near total in any conceivable realistic case, since $\omega h/c\ll1$. The large outward fluxes found by \citeauthor{Ler83aa} are an artefact of his prescribed driver $\xi_t(x,t)=V\,\rme^{-\rmi\omega t}$ at the base $x=0$ and the near-impermeability of the top setting up a weakly leaky resonant cavity. Then $\xi =\rmi\, V\omega^{-1}\,J_{-2\rmi\omega h/c}(r_0\, \rme^{-x/2h})/J_{-2\rmi\omega h/c}(r_0)$ where $r_0=2\omega h/a_0$, with the near-zeros of the (complex) denominator defining the resonances. The resonant cavity issue will arise again in Section \ref{prescbase}.

\subsection{General Method to Impose Radiation Boundary Condition at Infinity and Why the Exponential Atmosphere is Transparent}

Asymptotically flat profiles such as those considered above are ideally suited to the calculation of overall reflection coefficients, no matter how steep and non-WKB they may become at intermediate $x$. However, this does not mean that we can determine the reflectivity of a section of a non-WKB $a$ by abruptly sandwiching it between two uniform or WKB regions. Even a segment of the clearly non-reflective $a=x^2$ bookended continuously by two uniform sections will return non-zero reflection, but this is entirely due to discontinuities in $a'$.

For unbounded Alfv\'en profiles such as $a_1(x)$ though, there is no WKB region as $x\to+\infty$ in which inward and outward waves can easily be identified. However, the change of spatial variable to $r=\rme^{-x/2h}$ or similar that maps $x=+\infty$ to $r=0$ and $-\infty<x<+\infty$ to $r>0$ presents a convenient mathematical device for allowing complete absorption of a harmonic wave at $x=+\infty$. Equations such as (\ref{axisym}) and (\ref{apower}), which are symmetric in $r$,  suggest that waves be allowed to propagate through to negative $r$, thereby removing energy from the physical system. However, since the solutions generally contain a branch point at $r=0$ and a branch cut along the negative $r$-axis, it is necessary to treat $r$ as a complex variable and address the issue of whether to adopt the solution above or below the cut. The solutions (\ref{Hank2}) and (\ref{apowerHank2}) must in fact be continued below the cut to obtain the flow-through behaviour at $r=0$, based on the analytic continuation formula for Hankel functions, $H_\mu^{(2)}(z\, \rme^{-\rmi\pi}) = - \rme^{\rmi\mu\pi} H_\mu^{(1)}(z)$ \cite[formulae 10.11.5]
{DLMF}.\footnote{\new Alternatively, keep $r$ real but insert a weak frictional force [$-\gamma\,\partial\xi/\partial t$] on the right-hand side of Equation (\ref{axisym}), with $0<\gamma\ll\omega$. Then the upgoing solution is $H_0^{(2)}(\omega r\sqrt{1+i\,\gamma/\omega})$. This places the argument of the Hankel function below the cut for negative $r$.} This represents a mirror image outgoing wave as $r\to-\infty$, and is exactly the required energy sink. Both $H_\mu^{(1)}(z\, \rme^{-\rmi\pi})$ (below cut) and $H_\mu^{(2)}(z\, \rme^{\rmi\pi})$ (above cut) split into linear combinations of $H_\mu^{(1)}(z)$ and $H_\mu^{(2)}(z)$ functions \cite[10.11.8 and 10.11.4]{DLMF}
, breaking the symmetry and precluding a similar construction for solutions such as (\ref{smoothcap}) that arise for complex Frobenius index.

These considerations indicate that it is possible to mathematically prescribe a boundary condition that perfectly absorbs harmonic waves at $x=+\infty$. For Alfv\'en profiles with wave equation isomorphic to Bessel's equation with real order, the appropriate purely outgoing solution is simply the one involving $H_\mu^{(2)}(\omega r)$, \emph{i.e.}, $H_0^{(2)}(\omega r)$ for the exponential profile and $r^\mu H_{|\mu|}^{(2)}(\omega r)$ for the power-law profiles. In neither case is there any reflection. Essentially, this is a consequence of analyticity and symmetry in the complex plane.

For an initial-value problem though, the transient may not be perfectly absorbed. This is investigated next for the exponential profile.



\section{Exponential Alfv\'en Profile Initial Value Problem} \label{sec:ivp}
We now examine two initial value problems in the exponential atmosphere using the method of Laplace transformation. First we consider a harmonic point source at $r=R$ that switches on at time $t=0$, with either the outgoing radiation or the regularity boundary condition applied at $r=0$ ($x=+\infty$). Then we allow the point-source driver to switch off again at $t=T$. We are interested in the temporal dependence of the transient, and whether it is related to the reverberation (wake) seen in the steady state solution Equation (\ref{Hank2}). 

Our solution of the initial-value problem for Alfv\'en waves differs from that of \inlinecite{AnMusMoo89aa} in several ways: they solve the wave equation numerically rather than via Laplace transformation; they impose a base velocity whereas we impose a point force (acceleration); and most importantly they specify a totally reflective boundary condition at $x=+\infty$. Their solutions therefore reflect from infinity to set up a standing wave in finite time. We address this model too, but only to compare with the radiation boundary condition results.

\subsection{Point Source Harmonic Switch On} \label{switchon}
Consider the Alfv\'en wave equation with a point driver at $r=R$,
\begin{equation}
\pderivd{\xi}{t} = \frac{1}{r} \pderiv{}{r}\left( r \pderiv{\xi}{r} \right) + 8R^{-1}F(t)\,\delta(r-R)
\label{axisymF}
\end{equation}
where again $r = 2 h/a$, $F$ is an as yet arbitrary function of time that switches on at $t=0$, and a convenient normalization has been adopted. Laplace transforming Equation (\ref{axisym}) with initial conditions $\xi(r,0)=0$, $\xi_t(r,0)=0$ results in
\begin{equation}
\pderivd{u}{r}+\frac{1}{r}\pderiv{u}{r}+\omega^2 u = -8R^{-1}f(\omega)\,\delta(r-R)\,,
\end{equation}
where $u(r,\omega)=\int_0^\infty \rme^{\rmi\,\omega\,t} \xi(r,t)\,\rd t=\mathcal{L}\{\xi(r,t)\}$ and $f(\omega)=\mathcal{L}\{F(t)\}$. Imposing outgoing radiation boundary conditions at both $r=0$ and $\infty$, the Laplace transformed solution may be constructed as a Green's function,
\begin{equation}
u(r,\omega) =2\,\pi\,\rmi\,f(\omega)\,H_0^{(1)}(\omega r_>)\,H_0^{(2)}(\omega r_<)\,,
\end{equation}
where $r_<=\min(r,R)$ and $r_>=\max(r,R)$. The complex inversion formula then returns
\begin{equation}
\xi_\text{rad}(r,t) = \rmi\,\int_{-\infty+\rmi\varepsilon}^{\infty+\rmi\varepsilon} \rme^{-\rmi\omega t} f(\omega)\,
H_0^{(1)}(\omega r_>)\,H_0^{(2)}(\omega r_<)\,\rd\omega\,,
\label{ILT}
\end{equation}
where $\varepsilon$ is a positive constant sufficient to place the contour above any singularities in the complex $\omega$-plane. The subscript ``rad'' indicates that the radiation condition is applied at $r=0$.\footnote{All formulae used in this section relating to Bessel functions may be found in \inlinecite{DLMF}
, particularly the asymptotic formulae 10.17.5-6 and the analytic continuations 10.11.5 and 10.11.7-8. }

\begin{figure}
\begin{center}
\includegraphics[width=\textwidth]{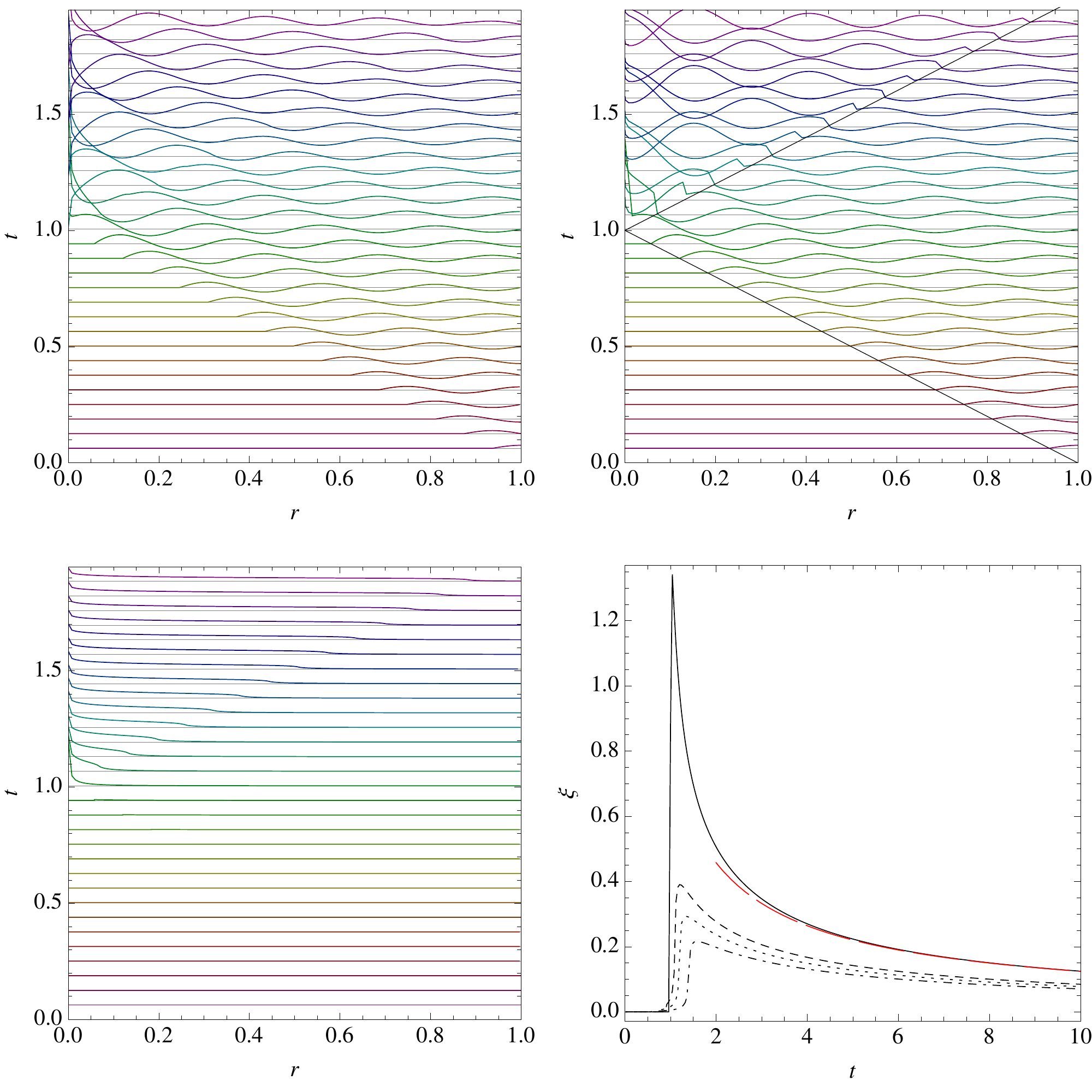}
\caption{Top row: Stack plots of $\re(\xi)$ against position $r$ and time $t=0.02\pi,\,0.04\pi, \ldots,\,0.6\pi$ for the case $R=1$ and $\Omega=25$. Left: $\xi_\text{rad}$; right: $\xi_\text{reg}$. Notice how in the left panel the phase propagates uniformly to the left throughout, corresponding to the outgoing radiation condition, whereas in the right frame a standing wave is progressively set up after the wave reaches $r=0$. The diagonal black lines in the right panel are $t=R\pm r$; reflection is evident only for $t>R+r$. Bottom left: Transient $\xi_\text{tr}$ for the case of the top-left panel. Bottom right: $\re(\xi_\text{tr}(r,t))$ for $r=0.002$ (full curve), 0.1 (dashed), 0.2 (dotted), and 0.4 (chained) against time [$t$], showing the slow monotonic decay of the transient. The long-dashed red curve represents the asymptotic formula (\ref{xitrAs}) to leading order $t^{-1}\ln t$ for the $r=0.002$ case.}
\label{fig:stack}
\end{center}
\end{figure}

Specializing now to the case of a monochromatic harmonic driver $F(t)=\rme^{-\rmi\Omega t}$ with $\Omega>0$, we have $f(\omega)=\rmi/(\omega-\Omega)$, and only require $\varepsilon>0$. With this choice of $F(t)$, $\re(\xi)$ corresponds to driver $-8R^{-1}\cos\Omega t$ and $\im(\xi)$ to $-8R^{-1}\sin\Omega t$.
The integral in Equation (\ref{ILT}) may be evaluated by completing the Bromwich contour in the upper half-plane (UHP) if $|r-R|>t$, and in the lower half-plane if $|r-R|<t$. The former yields $\xi_\text{rad}=0$ since there are no singularities in the UHP. This is expected as the signal from the source travels at unit speed in $r$-space. The $|r-R|<t$ case picks up contributions from the simple pole at $\omega=\Omega$, and from the branch cut conventionally lying along the negative real axis for Hankel functions. The pole produces the steady state and the branch cut contributes the transient: 
\begin{equation}
\begin{split}
\xi_\text{rad}(r,t) &=
-\,\int_{-\infty+\rmi\varepsilon}^{\infty+\rmi\varepsilon} \frac{\rme^{-i\omega t}}{\omega-\Omega}\,
H_0^{(1)}(\omega r_>)\,H_0^{(2)}(\omega r_<)\,\rd\omega   \\[6pt]
&=
\biggl\{
 \xi_\text{tr}(r,t) + 2\,\pi\,\rmi\,\rme^{-\rmi\Omega t} H_0^{(1)}(\Omega r_>)\,H_0^{(2)}(\Omega r_<)
\biggr\}\,
 \mathcal{U}(t-|R-r|)\,,
 \end{split}
\end{equation}
where $\mathcal{U}$ is the unit step function and the transient is
\begin{equation}
\begin{split}
\xi_\text{tr}(r,t) &= 2\,\int_0^\infty \frac{\rme^{\rmi xt}}{x+\Omega}\,
 \left[ H_0^{(1)}(x r)\,H_0^{(1)}(x R)-H_0^{(2)}(x r)\,H_0^{(2)}(x R) \right]\, \rd x \\[6pt]
&= 4\,\rmi \int_0^\infty \frac{\rme^{\rmi xt}}{x+\Omega}\,
\left[
J_0(xr)Y_0(xR)+J_0(xR)Y_0(xr)
\right]\,\rd x \,.
\end{split}
\label{transient}
\end{equation}

The oscillatory Fourier integral for $\xi_\text{tr}$ given in Equation (\ref{transient}) is readily evaluated numerically using Mathematica's built-in \textsf{NIntegrate} function. An alternative non-oscillatory formulation, valid only for $t>R+r$, is
\begin{multline}
\xi_\text{tr}(r,t) = \int_0^\infty \frac{8\,\rme^{-yt}}{\rmi\,\Omega-y}  \Bigl[
I_0(yr)I_0(yR)+\frac{\rmi}{\pi}\bigl(
K_0(yr)I_0(yR)+K_0(yR)I_0(yr)
\bigr)
\Bigr]\,\rd y\,.
\end{multline}
This is numerically simpler and quicker to evaluate, and more convenient for asymptotics. Specifically, we may use Laplace's method \cite{BenOrs78aa} to show that
\begin{equation}
\xi_\text{tr} \sim \frac{8}{\pi\,\Omega\, t} \left(-\rmi\,\pi+\ln\frac{4t^2}{rR}\right) 
- \frac{8}{\pi\,\Omega^2\, t^2} \left(-2\,\rmi + \pi+\rmi\,\ln\frac{4t^2}{rR}\right) 
\quad \text{for $t\gg R+r$,}   \label{xitrAs}
\end{equation}
correct to $\mathcal{O}(t^{-2})$.
It is interesting that the asymptotic decay of $\re(\xi_\text{tr})$ is monotonic, and depends only logarithmically on $r$ (linearly on $x$), whilst the decay of $\im(\xi_\text{tr})$ is independent of $r$ to leading order. {\new The logarithmic dependence on $r$ reflects the small-$r$ structure of the reverberation [$\beta$].}

\begin{figure}
\begin{center}
\includegraphics[width=\textwidth]{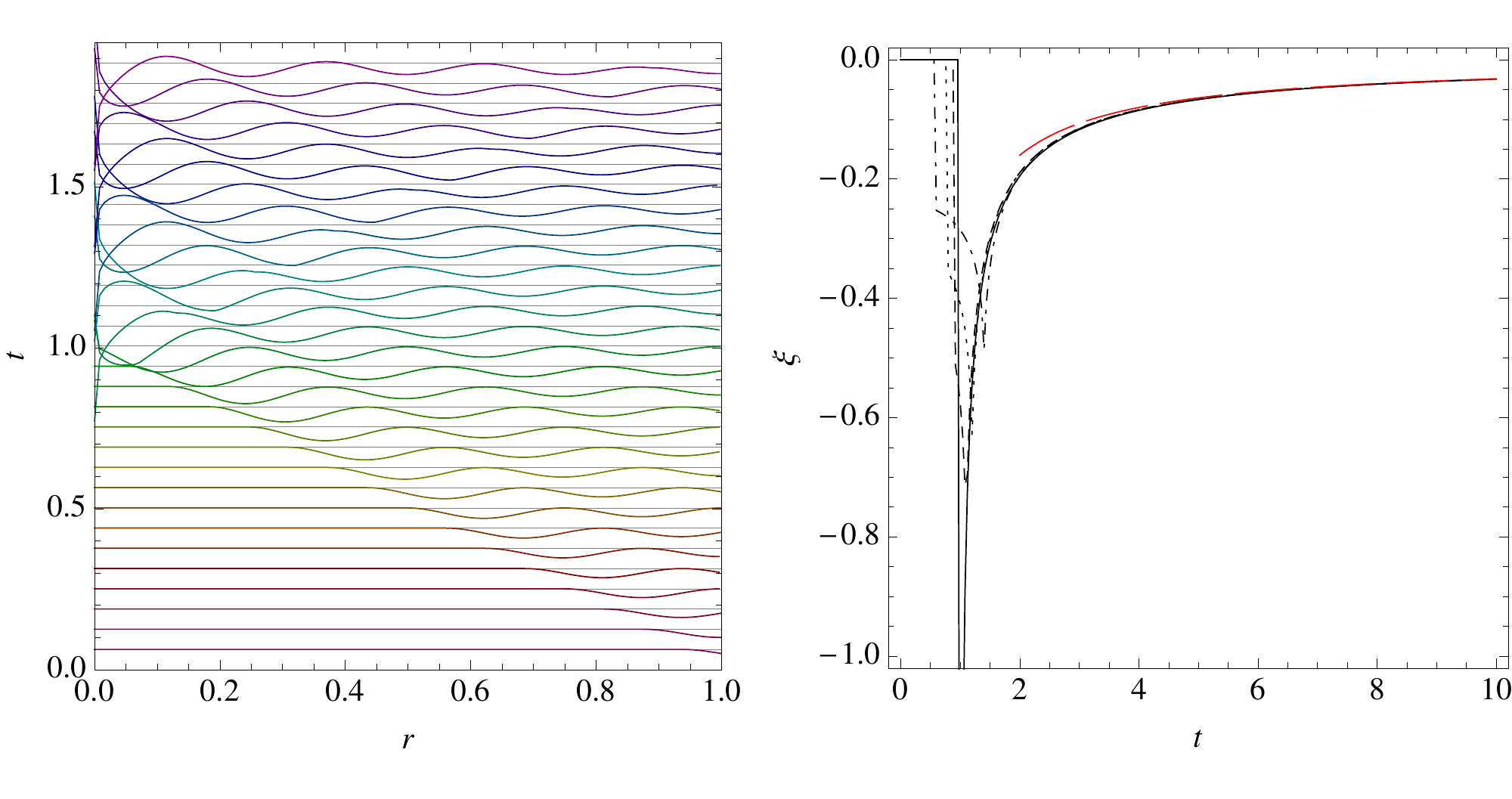}
\caption{Same as the top-left and bottom-right panels of Figure \ref{fig:stack}, but for $\im(\xi_\text{rad})$. In this case, as suggested by Equation (\ref{xitrAs}), the $t^{-1}$ decay is independent of $r$ to leading order.}
\label{fig:rowIm}
\end{center}
\end{figure}

For comparison, the same problem but with the usual regularity boundary condition applied at $r=0$ rather than the radiation condition yields
\begin{equation}
\xi_\text{reg}(r,t) =
-2\,\int_{-\infty+\rmi\varepsilon}^{\infty+\rmi\varepsilon} \frac{\rme^{-\rmi\omega t}}{\omega-\Omega}\,J_0(\omega r_<)\,
H_0^{(1)}(\omega r_>)\,\rd\omega \,.
\label{reg}
\end{equation}
A similar calculation was performed  by \inlinecite{BogCal97aa} for Alfv\'en waves in polytropic atmospheres with point impulse initial conditions and a regularity boundary condition where the density vanishes.
Clearly, since $2J_0(z)=H_0^{(1)}(z)+H_0^{(2)}(z)$,
\begin{equation}
 \Delta(r,t) \equiv \xi_\text{rad}(r,t) -\xi_\text{reg}(r,t) =
\int_{-\infty+\rmi\varepsilon}^{\infty+\rmi\varepsilon} \frac{\rme^{-\rmi\omega t}}{\omega-\Omega}\,H_0^{(1)}(\omega r_<)\,
H_0^{(1)}(\omega r_>)\,\rd\omega\,,
\label{Delta}
\end{equation}
which vanishes for $t<R+r$ since then the contour must be completed in the UHP. This confirms that the two solutions at $r$ are identical up until the time that the signal in the regular case reflects from the origin and returns to $r$. This is as it must be. Evaluating Equation (\ref{Delta}) by contour integration,
\begin{multline}
 \Delta(r,t) =-\mathcal{U}(t-r-R) \Bigl\{
 2\, \int_0^\infty \frac{\rme^{\rmi xt}}{x+\Omega}\, 
 \bigl(
 5J_0(xr)J_0(xR)-Y_0(xr)Y_0(xR)\\
 +\rmi\, [J_0(xr)Y_0(xR)+J_0(xR)Y_0(xr)]
 \bigr)\,\rd x
 +2\,  \pi\,\rmi\, \rme^{-\rmi\Omega t}\, H_0^{(1)}(\Omega r)\,H_0^{(1)}(\Omega R)
 \Bigr\}\,.
 \label{DeltaC}
\end{multline}

Figure \ref{fig:stack} (top row) illustrates the difference between the two solutions. They are indeed identical until $t=R$, but thereafter the reflected wave in the ``regular'' case progressively sets up a standing wave \cite{AnMusMoo89aa}. We also see that, even in the ``radiation'' case, a very slowly decaying transient is set up that spreads to progressively larger $r$ (bottom row). 

Figure \ref{fig:rowIm} displays the behaviour of $\im(\xi_\text{rad})$ and the asymptotic decay of its transient, which is now independent of $r$, as expected from Equation (\ref{xitrAs}).

We mention briefly that the present analysis may be replicated for the smoothly capped exponential profile (not presented here), although with added mathematical difficulty due to the frequency [$\omega$] appearing in both the argument and the order of the Bessel solutions. As expected, this case displays a marked reflection of the transient off the transition from exponential to flat Alfv\'en speed, but it also leaves a wake that decays as $\mathcal{O}(t^{-1})$.

\subsection{Point Source Harmonic Switch-On/Switch-Off} \label{switchonoff}
Now consider $F(t)=\rme^{-\rmi\Omega t}\,\mathcal{U}(T-t)$, where $T>0$ is a finite driving interval. Then $f(\omega)=\rmi(1-\rme^{(\omega-\Omega)T})/(\omega-\Omega)$ has only a removable singularity, so there is no residue and therefore no non-trivial steady state solution (of course). Adopting the radiation boundary condition again, the analysis is hardly changed and we find
\begin{equation}
\begin{split}
\xi_T(r,t) &=  -\,\int_{-\infty+\rmi\varepsilon}^{\infty+\rmi\varepsilon} \frac{\rme^{-\rmi\omega t}(1-\rme^{\rmi(\omega-\Omega)T})}{\omega-\Omega}\,
H_0^{(1)}(\omega r_>)\,H_0^{(2)}(\omega r_<)\,\rd\omega\\[6pt]
&= \xi_\text{rad}(r,t) - \rme^{-\rmi\Omega T}\,\xi_\text{rad}(r,t-T)\,.
\end{split}
\label{xiT}
\end{equation}
This is a simple consequence of the second shifting property of Laplace transforms \cite{Spi65aa}. The decay of the associated transient is therefore given asymptotically by
\begin{multline}
\xi_{T,\text{tr}} =
\frac{8(1-\rme^{-\rmi\Omega T})}{\pi\,\Omega\, t} \left(-\rmi\,\pi+\ln\frac{4t^2}{rR}\right)\\[6pt]
-\frac{8(1-\rme^{-\rmi\Omega T}-\rmi\,\Omega\,T)}{\pi\,\Omega^2\, t^2} \left(-2\rmi+\pi+\rmi\ln\frac{4t^2}{rR}\right)
 + \mathcal{O}(t^{-3}\ln t) \quad \text{for $t\gg R+r$.}
 \label{Ttrans}
\end{multline}
The decay is then $\mathcal{O}(t^{-2})$ or $\mathcal{O}(t^{-2}\ln t)$ if $\Omega T=2n\pi$ for integer $n$ since the two individual transients cancel to leading order. We shall refer to this case as ``balanced''. In the unbalanced case, the generic slow $\mathcal{O}(t^{-1})$ or $\mathcal{O}(t^{-1}\ln t)$ decay ensues.

\begin{figure}
\begin{center}
\includegraphics[width=\textwidth]{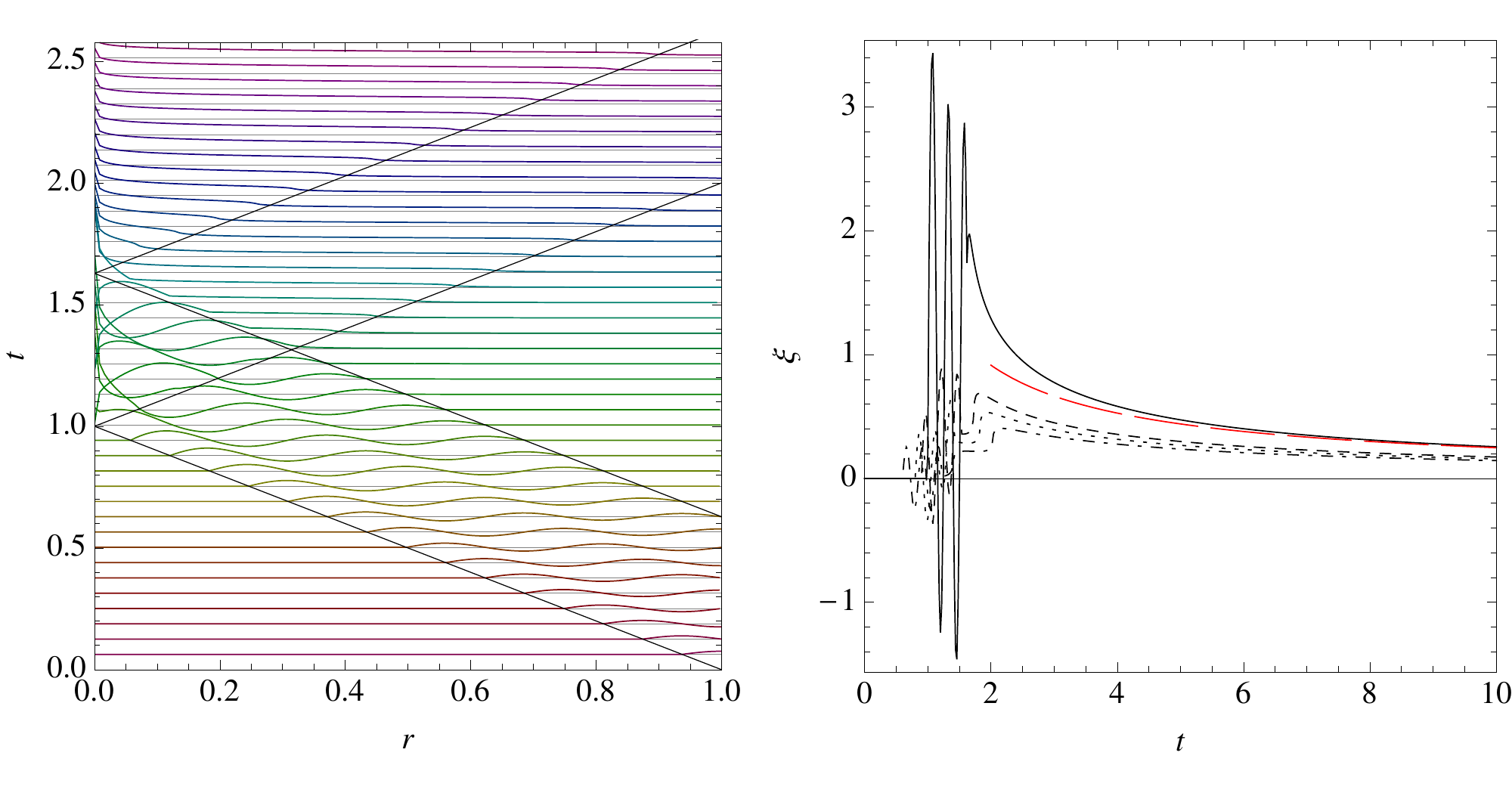}
\caption{Left: Stack plot of $\re(\xi_T)$ against position $r$ and time $t=0.02\pi,\,0.04\pi, \ldots,\,0.08\pi$ for the unbalanced case $R=1$, $\Omega=25$, and $T=\pi/5$.  The wake in the ``V'' at the right ($R-r+T<t<R+r$, $T/2<r<1$) is extremely weak and not discernible on the figure. It is non-zero though. Right: $\re(\xi_T(r,t))$ for $r=0.002$ (full curve), 0.1 (dashed), 0.2 (dotted), and 0.4 (chained) against time $t$. The red dashed curve depicts the leading order asymptotic transient formula from Equation (\ref{Ttrans}) at $r=0.002$.}
\label{fig:pT}
\end{center}
\end{figure}

\begin{figure}
\begin{center}
\includegraphics[width=\textwidth]{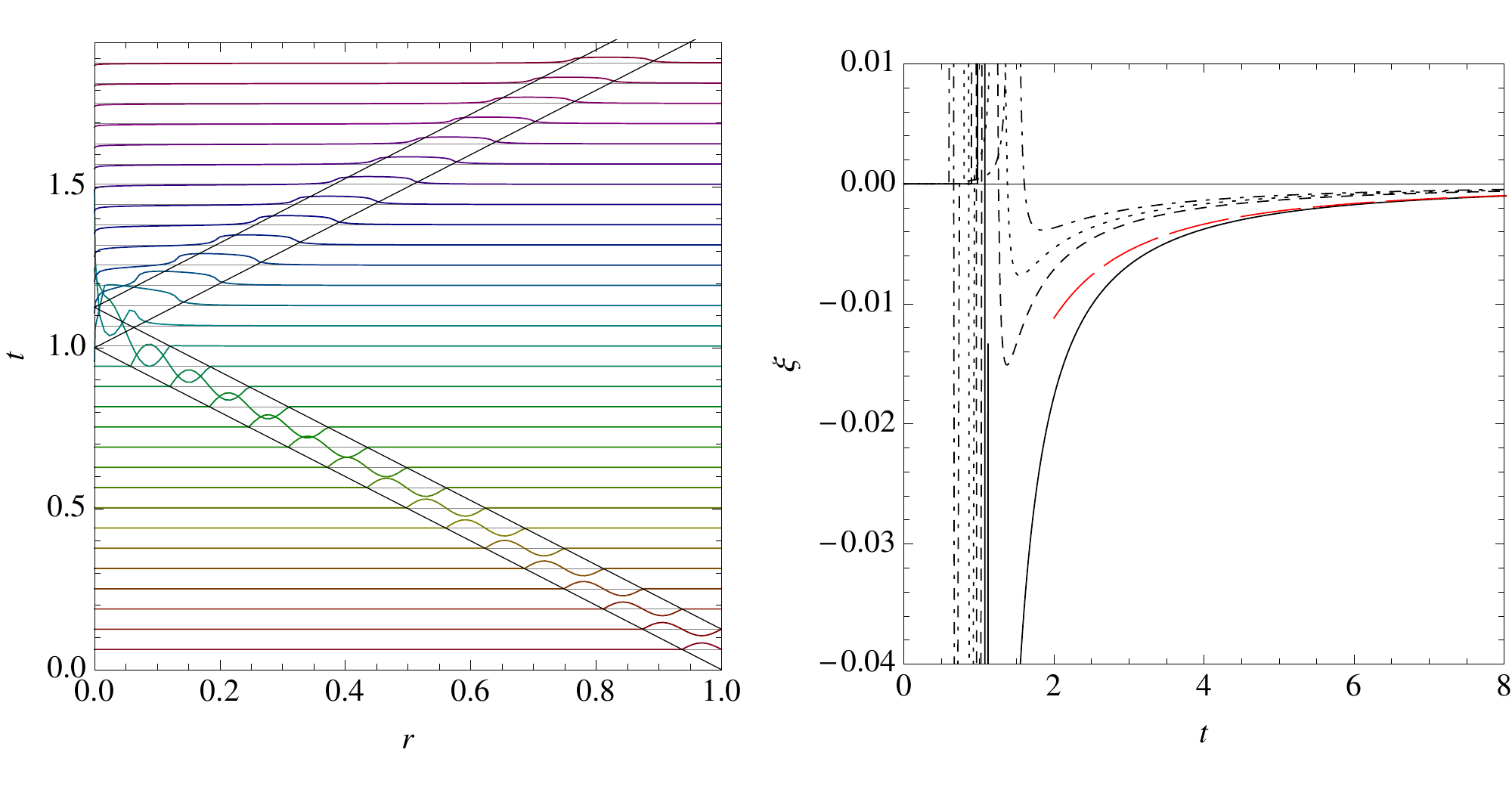}
\caption{Left: Stack plot of $\re(\xi_T)$ against position $r$ and time $t=0.02\pi,\,0.04\pi, \ldots,\,0.08\pi$ for the balanced case $R=1$, $\Omega=50$, and $T=\pi/25$.  Right: $\re(\xi_T(r,t))$ for $r=0.002$ (full curve), 0.1 (dashed), 0.2 (dotted), and 0.4 (chained) against time $t$. The red dashed curve depicts the leading-order asymptotic transient formula (\ref{Ttrans}) at $r=0.002$, which is now $\mathcal{O}(t^{-2}\ln t)$. Note also that the amplitude of the transient is far smaller than for the unbalanced case of Figure \ref{fig:pT}.}
\label{fig:pTS}
\end{center}
\end{figure}

Figure \ref{fig:pT} for an unbalanced case shows how the travelling wave signal moves leftward and out of the domain, but leaves a rightward-spreading and slowly decaying transient long after the driver has been switched off. Figure \ref{fig:pTS} depicts a balanced case where very little wake is left at large $t$. Nevertheless, although the oscillatory (steady state) signal has escaped the domain at $r=0$, a non-oscillatory transient is reflected in a narrow band in $r$--$t$ space that represents the absence of balance in the initial switch-on. It is soon cancelled by the subsequent symmetric switch-off.

\subsection{Prescribed Base Harmonic Displacement}\label{prescbase}

For completeness, we briefly mention the case where the base ($r=R$) velocity or displacement is prescribed rather than introducing a driving force term as above. This is the more common procedure {\new \cite{Hol78aa,Ler83aa,SchCalBel84aa,AnMusMoo89aa}}, and leads to probably unrealistic Alfv\'en-frequency resonances because the ``closed box'' has natural frequencies. 

\begin{figure}
\begin{center}
\includegraphics[width=\textwidth]{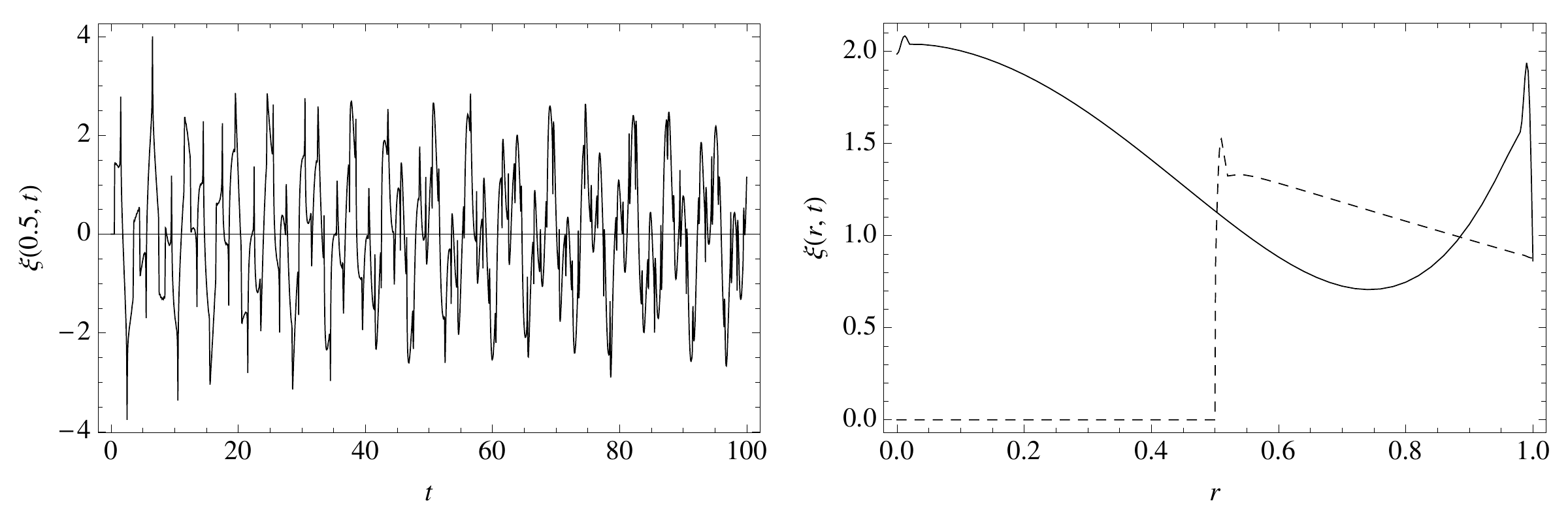}
\caption{Left: Temporal evolution of $\re\xi$ at $r=0.5$ for the rigid harmonic driver case $\Omega=R=1$ of Equation (\ref{xiBox}). The first 50 Bessel zeros are retained. The displacement switches on at $t=R-r=0.5$ and oscillates wildly due to the many progressively out of phase trapped Bessel modes. Right: $\re\xi$ \emph{vs.}~$r$ for the same case, at times $t=100$ (full) and $t=0.5$ (dashed), displaying smooth behaviour in space. For this plot, 100 Bessel modes were used.}
\label{fig:tBox}
\end{center}
\end{figure}

Specifically, letting $\xi(R,t)=\rme^{-\rmi\Omega t}$ and adopting the regularity boundary condition at $r=0$, the solution assuming zero initial displacement and velocity is obtained by Laplace inversion:
\begin{equation}
\begin{split}
\xi(r,t) &= \frac{\rmi}{2\pi} \int_{-\infty+\rmi\varepsilon}^{\infty+\rmi\varepsilon}
\frac{\rme^{-\rmi\omega t}}{\omega-\Omega}\, \frac{J_0(\omega r)}{J_0(\omega R)}\,\rd\omega \\[6pt]
&=\left\{
\rme^{-\rmi\Omega t}\frac{J_0(\Omega r)}{J_0(\Omega R)}
-\sum_{n=0}^\infty \frac{J_0(j_nr/R)}{J_1(j_n)} 
\left[ \frac{\rme^{-\rmi j_nt/R}}{j_n-\Omega R} +  \frac{\rme^{\rmi j_nt/R}}{j_n+\Omega R}\right]
\right\}\mathcal{U}(t+r-R)\\[4pt]
\end{split}
\label{xiBox}
\end{equation}
where the $j_n$ ($n=1$, 2, \ldots) are the positive zeros of $J_0$, and $0<r\le R$. Asymptotically, the $j_n$ are nearly uniformly distributed: $j_n\sim a_n+1/(8a_n)$ as $n\to\infty$, where $a_n=(n-\frac{1}{4})\pi$ \cite[formula 10.21.19]{DLMF}
, so simple poles are distributed along the whole real line, specifically at $\omega=\pm j_n/R$. In deriving this solution it has been assumed that $\Omega\ne j_n$ for any $n$. The solution for the transient is no longer as neatly expressed as before, since there are now no branch points and hence there is no continuous spectrum and no integral associated with a branch cut. This is because the transient cannot escape the box, either through $r=0$ or $r=R$. Each initially excited mode $n$ persists indefinitely as a discrete harmonic of the closed cavity $0<r<R$, resulting in a chaotic time evolution but smooth spatial structure (Figure \ref{fig:tBox}). This is a consequence of treating the bottom boundary as rigid for all frequencies other than the driving frequency $\Omega$. In reality, Alfv\'en waves should be able to penetrate into the solar interior due to their large wavenumber in the low atmosphere.

The case with radiation boundary condition at $r=0$ is more complex again, as branch points are involved as well as the complex zeros of $H_0^{(2)}$. These complications are incidental to our main concerns though, and will not be pursued further here.

\section{Summary and Discussion}
We recapitulate our main findings.
\renewcommand{\labelenumi}{\roman{enumi})}
\begin{enumerate}
\item Reflection of time-harmonic Alfv\'en waves vanishes in the high frequency (WKB) limit provided the Alfv\'en speed profile $a(x)$ is continuous. If $a$ is a $\mathscr{C}^{d-1}$ function, \emph{i.e.} $(d-1)$-times continuously differentiable but not more, then the reflection coefficient $\mathcal{R}=\mathcal{O}(\omega^{-2d})$ as $\omega\to\infty$. If $a$ is a $\mathscr{C}^\infty$ function, then $\mathcal{R}\to0$  exponentially as $\omega\to\infty$.
\item There exist atmospheres, $a=a_0 x^{(n-1)/(n-2)}$ for integer $n$, where Alfv\'en wave propagation is isomorphic under the transformation $r=(n-2)\,a_0^{-1}x^{-1/(n-2)}$ to the $n$-spherically symmetric solutions of the $n$-dimensional wave equation with unit wave speed in $r$-space. Only for odd-integer $n$ do disturbances travel without trailing a wake. The formulation of the Alfv\'en wave equation as an $n$-dimensional spherical wave equation in a homogeneous medium provides a simple explanation of the Alfv\'en wakes found by \inlinecite{HolIse07aa} using a characteristics and impulse-function approach, and the ``continual coupling'' inferred by \inlinecite{Ler80aa} using matrix diagonalization.
\item For $n$ not an odd integer, time-harmonic radiation solutions for these power-law atmospheres, and the exponential atmosphere which corresponds to $n=2$, contain branch points and therefore exhibit the characteristics of a continuous spectrum, including algebraic time decay of transients.
\item Another feature of these solutions is wakes. Wakes may be thought of as a form of (continuous) reflection, but are distinct from simple reflection which sends a signal back from whence it came. The existence, or otherwise, of branch points is a feature of the differential equation, and can be determined by local analysis about its singular points without needing to solve the DE.
\item The atmosphere $a=a_1=a_0\,\rme^{-x/2h}$ corresponds to the case $n=2$ under the transformation $r=2h/a$, and so does not admit relatively undistorted solutions. The initial-value problem therefore produces a wake that persists even after the driver is switched off. However, the steady-state component consists of a unidirectional travelling part and a stationary reverberation. There is no reflection in the standard sense. With the radiation boundary condition applied at $r=0$ ($x=+\infty$) the steady-state signal freely leaves the domain, but a weak non-oscillatory transient remains that decays as $\mathcal{O}(t^{-1})$ or $\mathcal{O}(t^{-1}\ln t)$ (unbalanced case), or $\mathcal{O}(t^{-2})$ or $\mathcal{O}(t^{-2}\ln t)$ (balanced case), depending on the form of the driver. This is a sort of slow monotonic relaxation to the equilibrium state. Although the steady-state solution suffers no reflection at $r=0$, the transient does reflect, spreading backwards (ever weakening) at unit speed.
\item Partial reflection can be associated with complex Frobenius indices at $r=0$. For real indices it is possible to construct a radiation boundary condition at $x=+\infty$ based on the analyticity and symmetry of the wave equation in $r$ that yields no reflection of harmonic waves at any frequency.
\end{enumerate}

\section{Conclusions}
The analysis presented here may seem esoteric. Nevertheless, it has practical implications for the way that we model wave propagation in the solar atmosphere. There are any number of articles \cite[for example]{LeeHolFla82aa} that seek to calculate the ``intrinsic'' reflectivity of coronal Alfv\'en waves by placing a uniform slab above an exponential or similar model.  However, this says more about the matching point than about the underlying atmosphere. The exponential and power-law profiles are cases in point. If the atmospheres are allowed to extend unimpeded to infinity without truncation, they are entirely transparent. Or, more physically, if an efficient and non-reflective wave energy sink is placed high in the atmosphere, there is no reflection from the body of the atmosphere either. A simple uniform or WKB slab does not represent such a non-reflective sink, although this is not to say that it may not be a reasonable representation of the outer corona \cite{Ler81aa}. In the absence of any Alfv\'en wave dissipation, such a WKB top would indeed induce strong reflection in the underlying atmosphere. But the point is that the underlying atmosphere is not necessarily \emph{intrinsically} reflective (\emph{viz.}, the exponential or power-law Alfv\'en-speed profiles). It is the \emph{transition} to the uniform or WKB top that reflects.

This is relevant when performing numerical experiments in truncated model atmospheres. If we \emph{choose} to, we can postulate a radiation boundary condition at the top of our region of interest. {\new In numerical simulations, this is typically done using characteristic boundary conditions \cite{EngMaj77aa} or absorbing layers \cite{Ber94aa}. For steady monochromatic waves, the task may be easier, through simple matching to a known analytic solution that represents an outgoing wave.} For the exponential atmosphere, this involves adopting the $H_0^{(2)}(\omega r)$ Hankel function solution {\new\cite{CalGoo08aa}} . The $J_0(\omega r)$ solution on the other hand is appropriate if we want a reflective boundary {\new at infinity}. With this in mind, matching numerical solutions in a finite domain to a Hankel function or similar radiation solution is mathematically well founded and physically interesting. {\new This does not imply that the exponential atmosphere really does extend to infinity; it is simply an effective mathematical device for imposing a radiation condition at the top of a computational domain. A more realistic treatment of solar coronal Alfv\'en waves should in fact address the maximum and gradual decline in the Alfv\'en speed beyond a few solar radii and the loss of hydrostatic equilibrium that results in the solar wind \cite{Vel93aa}, but that is beyond the expository scope of the present analysis. It is notable though that \citeauthor{Vel93aa} finds near-total \emph{transmission} at high frequencies in this model, which further supports the use of a radiation boundary condition in simpler atmospheres.}

Analysis of the initial-value problem for the exponential atmosphere verifies that the steady state $H_0^{(2)}(\omega r)$ solution does indeed fully depart the physical model through $x=+\infty$, though a weak transient remains that decays algebraically in time. Since we might expect episodic generation of Alfv\'en waves in the lower atmosphere, for example by fast-wave conversion \cite{HanCal12aa}, the ubiquitous presence of such slowly relaxing transients can hardly be avoided. These motions would not be identified as waves observationally, as they are not oscillatory in time.


\bibliographystyle{spr-mp-sola}        
\bibliography{fred}

\end{document}